\documentstyle[epsfig]{mn}

\title[Constraints on dark matter physics from dwarf 
 galaxies through galaxy cluster haloes]
{Constraints on dark matter physics from dwarf
 galaxies through galaxy cluster haloes}

\author[Firmani, D'Onghia, Chincarini, Hern\'{a}ndez, Avila-Reese] 
{Firmani C.$^{1,5}$, D'Onghia E.$^{2}$, Chincarini G.$^{1,3}$, 
Hern\'{a}ndez X.$^{4}$, Avila-Reese V.$^{5}$\\
$^{1}$ Osservatorio Astronomico di Brera,
 via E. Bianchi 46, 23807 Merate (LC), Italy\\
$^{2}$ Universit\'{a} degli Studi di Milano, via Celoria 16,
 20100 Milano, Italy\\
$^{3}$ Universit\'{a} degli Studi di Milano-Bicocca, Italy\\
$^{4}$ Osservatorio Astrofisico di Arcetri, Largo E. Fermi 5,
 50125 Firenze, Italy \\
$^{5}$ Instituto de Astronom{\'\i}a, UNAM, A.P. 70-264, 04510
M\'{e}xico D.F., M\'{e}xico \\
E--mail: {\tt firmani@merate.mi.astro.it}, {\tt elena@merate.mi.astro.it},
{\tt guido@merate.mi.astro.it},\\ {\tt xavier@arcetri.astro.it},
{\tt avila@astroscu.unam.mx}
}

\date{\underline{submitted to MNRAS, 2000 March 22}}

\begin{document}
\maketitle

\begin{abstract}

One of the predictions of the standard cold dark matter 
model is that dark haloes have centrally divergent density
profiles. An extensive body
of rotation curve observations of dwarf and low surface
brightness galaxies shows
the dark haloes of those systems to be characterized by soft
constant density central cores. 
Several physical processes have been proposed to produce soft
cores in dark haloes, each one with different scaling properties.
With the aim of discriminating among them 
 we have examined 
the rotation curves of dark matter dominated
dwarf and low surface brightness galaxies and the inner mass profiles 
of two clusters of galaxies lacking a central cD galaxy
and with evidence of  soft cores in the centre. 
The core radii and central densities 
of these haloes scale in a well defined manner with the depth of their
potential wells, as measured through the maximum circular velocity.
As a result of our analysis we identify self-interacting cold dark matter
as a viable solution to the core problem, where 
a non-singular isothermal core is formed in the
halo center surrounded by a Navarro, Frenk, \& White profile in 
the outer parts.
We show that this particular physical situation 
predicts core radii in agreement with observations. 
Furthermore, using the observed scalings, we derive
an expression for the minimum 
cross section ($\sigma$) which has an explicit dependence 
with the halo dispersion
velocity ($v$). If $m_x$ is the mass of the dark matter particle:
$\sigma / m_x \approx 4 \ 10^{-25} (100 \ km s^{-1}/v)$ $cm^2/Gev$.   

\end{abstract}

\section{Introduction}
The predictions of the cold dark matter (CDM) models successfully 
account for several 
observational facts such as the distribution of matter on large scales, 
the uniformity of the cosmic microwave radiation and its small 
temperature anisotropies, and the measured values of the cosmological 
parameters. 
However, on small scales the predictions seem to be in conflict with 
observations suggesting that modifications 
to the standard scenario should be introduced. 
One of the problems of the CDM theory is that the 
inner density profile of the virialized haloes is too steep with 
respect to what is inferred from  rotation curves of dwarf spiral 
galaxies (Moore 1994; Flores \& Primack 1994; Burkert 1995). In fact it is 
well known that these galaxies are
systems strongly dominated by dark matter, so that their 
rotation curves are good tracers of the dark halo gravitational 
potential. A similar situation is expected for low surface brightness 
(LSB) galaxies (de Blok \& McGaugh 1997), (although the observational
evidence of a soft core in these cases has been challenged by some authors
 e.g., van den Bosch et al 1999; 
Swaters, Madore $\&$ Trewhella 2000). Similarly, 
Hern\'{a}ndez $\&$ Gilmore (1998) showed that 
the HI rotation curves of both LSB and normal large galaxies can be 
characterized by a significant soft core inner region. 
 The Tully-Fisher relation can be also better predicted by
galaxy formation models when a density profile shallower than
the CDM one is used (Firmani $\&$ Avila-Reese 2000; Mo $\&$ Mao 2000).
On the galaxy
cluster scale the presence of shallow halo cores is highly 
uncertain because   
of the small amount of data available. However,
from strong lensing observations of the cluster
CL0024+1654, Tyson, Kochansky $\&$ Dell'Antonio (1998) making  
a mass map of unprecedented resolution 
have revealed the presence of a 
soft core in conflict with the predictions of numerical simulations.

Recently several attempts have been made in order
to produce soft cores by a manipulation of the power spectrum of 
fluctuations. 
Moore and co-workers (1999b) have tested the effects
of introducing a lower cut-off in the power spectrum 
leading to a halo formation scenario tending towards a monolithic 
collapse, obtaining final density 
profiles actually steeper than the one found by Navarro, Frenk, \& White 
(1997, hereafter NFW). However, a lower cut-off
in the power spectrum with  presence  of some tangential
velocity dispersion  
  is able to produce
haloes with shallow cores (Avila-Reese, Firmani $\&$ Hern\'{a}ndez 1998). 
Hogan \& Dalcanton (2000) introduced the notion of a limiting
phase-space density for dark matter as a way to
restrict the volume density dark matter particles can attain.
An alternative scenario by  Peebles (2000) proposes a scalar field
dark matter treated
as an ideal fluid, in which a suppression of the small scales
in the power spectrum is an intrinsic property of the theory.
In addition, a model of repulsive dark matter has recently
been discussed by Goodman (2000).
Spergel \& Steinhardt (2000) have 
proposed  collisional CDM as an alternative solution
to soft core formation.   Several papers have appeared
investigating this idea (Hannestad 1999)
and the relevance of its astronomical implications (Ostriker 1999).
The case in which a very large particle scattering 
cross-section is assumed was studied within the hierarchical formation
picture using fluid dynamical techniques (Moore et al. 2000; Yoshida et
al. 2000). In this case dark matter behaves like a non-
dissipative fluid that can shock heat producing haloes with singular 
isothermal cores. If a weaker self-interaction is working between
particles during the merger history of the haloes, then core 
expansion due to a gravothermal instability is expected, as shown
with a thermodynamical approach by Firmani and co-authors (2000).
Numerical N-body simulations modified to investigate a weak self-interaction
have also shown this effect (Burkert 2000).

Another controversial question regarding standard hierarchical
structure formation models is that the predicted abundance of
galactic subhaloes is an order of magnitude higher than the
number of satellites actually observed within the Local Group
(Klypin  et al. 1999; Moore  et al. 1999a). 
A substantial advance in understanding of this question
was made by Moore  et al. (2000) and Yoshida et al. (2000). 
They find that, due to self-interaction, the predicted galactic
subhalo number decreases, even if in Yoshida et al. (2000)
the decrease appears rather moderate. Interestingly, Moore
 et al. (2000) point out that in a cosmological picture
in which self-interaction is working, ram-pressure could be more 
effective in producing stripping and viscous drag than tidal
forces and dynamical friction. However, treating self-interacting CDM as 
a fluid with very strong cross section these authors never find
soft cores.
A more conservative solution to the satellite question,
but unrelated to the core problem 
has been suggested by Bullock et al. (2000).

Given the multiplicity of proposed solutions that have appeared in 
answer to the core question, we carry out an analysis 
of the available observational data, in order to identify the 
proper mechanism for producing soft cores.
Taking into account the comparison with observations we conclude that
self-interacting CDM appears the most viable solution to produce soft
cores. We  point out 
 two limiting cases between
which dark matter self-interaction may generate isothermal cores:
when the cross section is sufficiently large to thermalize the entire halo,
and when a
minimum cross section allows self-interaction to yield
thermal equilibrium only in the core region.
In the following section we present  
the analysis of the observational
data, in section 3 we show the density profiles derived from CDM models
incorporating manipulations of the power spectrum of the primeval 
fluctuation field. In section 4 the    
implications of collisional dark matter are investigated, for the 
strong and minimum cross section cases. Our conclusions are presented
in section 5.
The cosmological model we use is flat with matter density $\Omega_m =0.3$,
cosmological constant $\Omega_{\Lambda}=0.7$ and expansion rate
$H_0=60$ km s$^{-1}$  Mpc$^{-1}$ (Saha et al. 1999). 

\section{Halo core scaling laws from observations}

Analysis of the inner halo profile is difficult for most galaxies
because of the ambiguities in the estimate of the stellar mass-to-light 
($M/L$) ratio and the resulting dynamical contribution of the stellar 
component to the observed rotation velocities. A further complication is
due to the fact that 
the halo inner density profile is affected by 
the gravitational pull of the baryonic matter. For these reasons, the
best candidates to study galactic dark haloes are dwarf and LSB galaxies,
i.e. galaxies which are dark matter dominated. The rotation curves of LSB
 galaxies  derived from  radio observations are rather uncertain in the
innermost region because of beam smearing  (van 
den Bosch et al. 1999). High-resolution rotation curves, however,  
have been obtained
in $H\alpha$ for five LSB galaxies  (Swaters et al. 2000). These
rotation curves 
rise more steeply in the inner parts than the ones obtained
by HI observations.
Our analysis is based on a sample of galaxies (i) 
strongly dominated by dark matter 
and (ii) with accurately measured rotation curves. The sample
consists of six dwarf galaxies, nine LSB galaxies, and two low-luminosity
disk galaxies (see Table 1 and the references therein).
On galaxy cluster scale, we shall use the mass 
distribution for the cluster CL0024+1654 derived from strong lensing 
techniques (Tyson et al. 1998), and for the cluster CL0016+16 derived
from weak lensing studies (Smail et al. 1995). In the former case, the 
spatial resolution attained is high, while in
the latter case the observational uncertainty in the central regions is 
very large. In both cases the inner mass distribution is soft,
there is no a  
massive cD galaxy and 
 one can assume that both clusters are dark matter dominated. 

We shall infer from
observations the behaviour of the halo core 
radius $R_c$ and of the 
central density $\rho_c$ with the halo maximum circular velocity $V_m$ (the 
gravitational potential well). All the data we use have been taken 
from the literature but the parameters we adopt have been estimated  
from our analysis.

\begin{figure*}
\epsfig{file=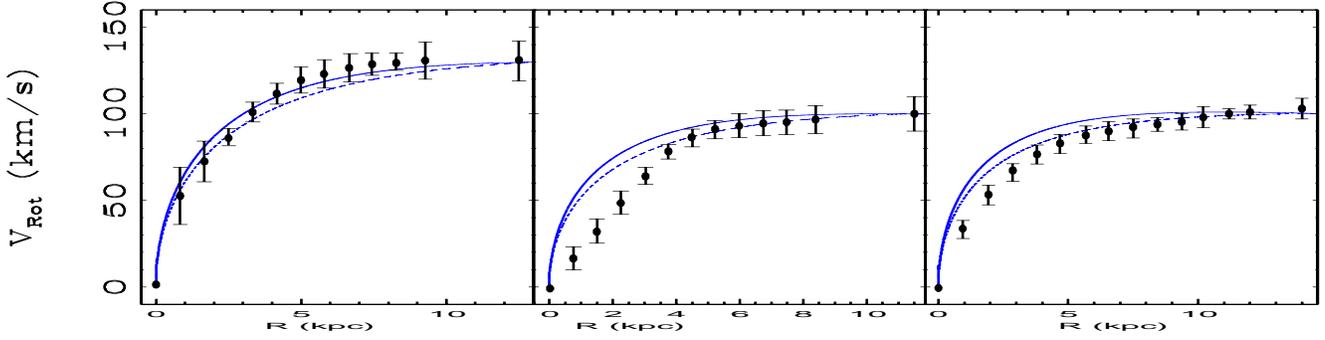,angle=0,width=19.8cm,height=4.96cm}
\@ 
\caption{Comparison of  rotation curves from 
Swaters et al. (2000) to
NFW profiles having the observed asymptotic velocity.  Without considering
disk formation processes, dashed curves,
and including disk formation processes, solid curves, for galaxies
F568-1, F568-3 and F574-1, in panels 1, 2 and 3, respectively.}
\end{figure*} 

\begin{figure*}
\epsfig{file=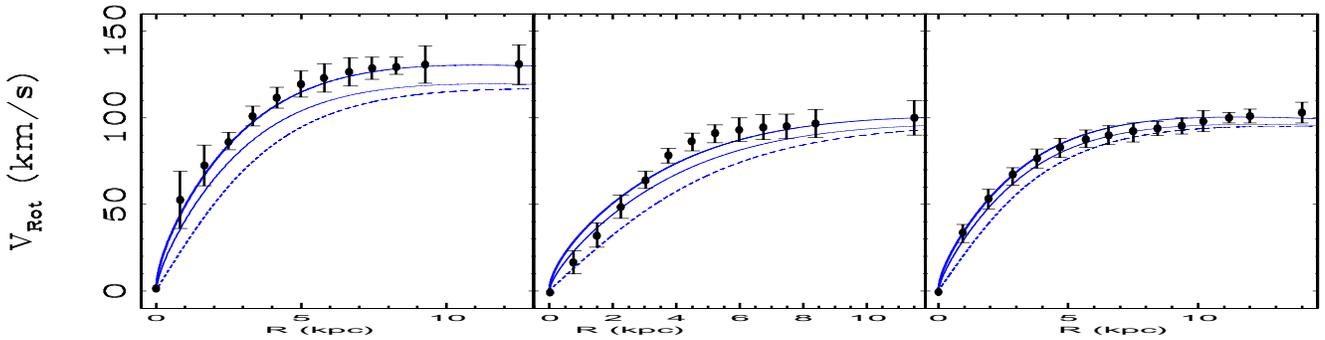,angle=0,width=19.8cm,height=4.96cm}
\@ 
\caption{Comparison of rotation curves from 
Swaters et al. (2000) to
King profiles, including disk formation processes. The  dashed lines
give the rotation curves of the initial dark halo; the thin solid ones 
give the rotation curves of the contracted dark halo after disk
formation and the final total rotation curve is shown by the thick solid
curves, for galaxies F568-1, F568-3 and F574-1, in panels 1, 2 and 3,
respectively.}
\end{figure*} 

By fitting the observed rotation curves
of the sample galaxies after subtraction of the disk contribution 
to the rotation 
curve  allows us to fit the two parameters
of a non-singular isothermal model, $R_c$ and $v$, where $v$ is 
the one-dimensional
velocity dispersion. The central density is then:
$\rho_c =9 \ v^2 /4 \pi G R_c^2$.
The properties of 
the sample objects are listed in Table 1. In the various columns 
we have: 
 the name, the maximum circular velocity $V_{m}$ in km/s,
the core radius $R_c$ 
in kpc, the 
central density $\rho_c$ in $M_{\odot}$/pc$^3$, and the references used. 

\begin{figure}
\epsfig{file=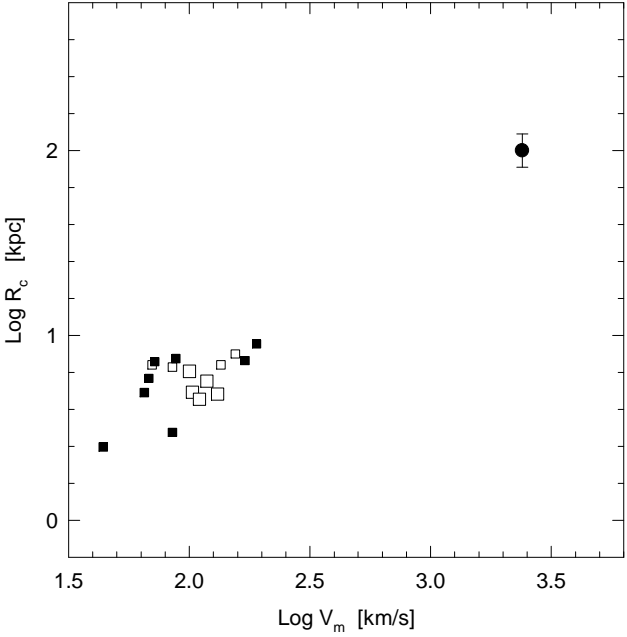,angle=0,width=\hsize,bbllx=79pt,bblly=225pt,
bburx=492pt, bbury=663pt, clip=} 
\@ 
\caption{The core radius vs. maximum circular velocity for dwarfs
(filled squares), LSB galaxies (empty squares) and CL0024+1654 cluster
(filled circle) listed in Table 1. 
The big empty square symbol is for LSBs by $H\alpha$
rotation curves (Swaters 2000).}
\end{figure}

\begin{figure}
\epsfig{file=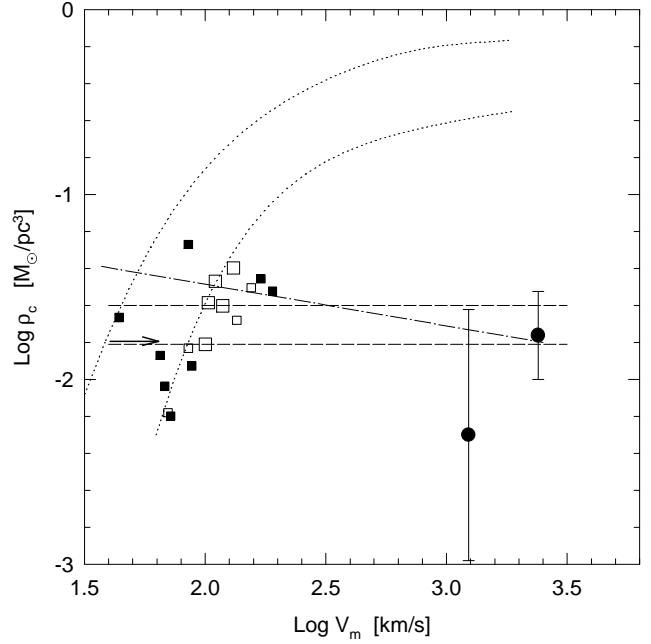,angle=0,width=\hsize,bbllx=79pt,bblly=225pt,
bburx=492pt, bbury=663pt, clip=}
\@ 
\caption{The halo central density vs. maximum circular velocity for dwarfs
(filled squares), LSB galaxies (empty squares) and galaxy clusters
(filled circle) listed in Table 1. Dotted curves show predictions for haloes
resulting from a lower cut-off in the power spectrum at 
$M=3 \ 10^9 \ M_{\odot}$ (upper dotted curve) and at 
$M=3 \ 10^{10} \ M_{\odot}$ (lower dotted curve) with  $\epsilon=0.27$
see text.  
Long-dashed lines are for halo central densities expected if 
 the halo mass fraction
accumulated up to $z=5$ (upper dashed line) or $z=4$ (lower dashed 
curve) in a CDM mass aggregation history actually
collapses at these redshift with  $\epsilon=0.27$. 
The point-dashed line represents the state
of maximum entropy predicted by a King profile 
 when a thermal equilibrium is reached.}
\end{figure}  

It is important to realize that a present day rotation curve
is not only the result of an initial dark halo structure, but is a 
consequence of the
process of disk formation, which as it will be shown, can alter
the inner shapes of galactic rotation curves.
Two effects contribute to
changing the density distribution of a dark halo during disk formation, both
relevant only in the central regions, precisely where the problem of the 
existence or absence of a soft core arises. Firstly, as a result of disk
formation, the baryonic component contraction makes the baryons in the
central regions approximately as dynamically relevant as the dark matter.
 Secondly, the contraction of the baryons alters the total mass
distribution in the inner regions, forcing the dark halo to adjust to the now
modified gravitational potential. 
These two points are essentially well known, and have been treated in standard
disk galaxy formation models, e.g. Fall \& Efstathiou 1980, Flores et al. 1993;
Dalcanton et al. 1997; Firmani \& Avila-Reese 2000 and
Hernandez \& Gilmore 1998  
who show explicitly the relevance of the above effects even in the central 
regions of LSB systems. 
A detailed modeling of the sample of five galaxies by 
  Swaters et al. (2000) allows us to quantify this possibility, and
to asses the evidence supporting shallow cores at this region of the
LSB galactic distribution. With such a modeling we can also estimate
the uncertainties of our other data.
For each of these  galaxies we use a simple disk 
formation model where a fraction $F$ of the original
halo mass is turned into an exponential disk 
having the disk scale length of the
observed galaxy. The gravitational pull suffered by the inner 
regions of the initial
dark halo is calculated using an hypothesis of adiabatic invariance  for the
orbits of the dark particles, and the final rotation curve calculated taking
into account the modified halo and the contribution of the exponential disk
(Hern\'{a}ndez $\&$ Gilmore 1998). 
 The parameters of this disk formation model were optimized 
using the full observed rotation curves, together with an imposed 
Tully-Fisher relation (Firmani $\&$ Avila-Reese 2000) and the 
observed disk scale radius of the galaxies in question. In this way
we solved for the values of F which best yield a final rotation
curve.
We have modeled these 
galaxies using two different dark halo structures,
 namely a NFW profile and a King profile.  
 The analytic NFW profile is $\rho(r)=\rho_s(r/r_s)^{-1}(1+r/r_s)^{-2}$
with $r_s$ the scale radius corresponding to the scale at which 
the logarithmic slope is -2. Since $\rho_c$ and $r_s$ are correlated 
and they depend only on the mass, the density profiles of haloes can
be described by a one-parameter family of models, where the 
parameter is the mass.
The King density profile has the form: $\rho=\rho_0 f(r/r_0,P)$, with
$\rho_0$ the central density and $r_0$ the core radius (the radius 
at which the density is roughly $\rho_0/3$). The form parameter is 
$P=\Psi(0)/\sigma^2$, being $\Psi(0)$ the relative gravitational
potential and $\sigma$ the linear velocity dispersion in the center.
We adopt the same formalism of Binney $\&$ Tremaine assuming for f the
same function displayed in their Fig.4-9.

Fig. 1 shows the rotation curves of 
F568-1, F568-3 and F574-1, in the 1st, 2nd and 3rd panels, respectively 
(dots with
error bars). The dashed curves give the
unique CDM NFW halo which has a maximum velocity matching that of the
observations, for each galaxy. Panel (1) refers to that galaxy from
Swaters et al. (2000) for which
the NFW profile fits best. Panel (2) gives the
opposite case, where the NFW profile fails completely to reproduce the
data. For the remaining three galaxies only marginal agreement between the
NFW model and the rotation curves is possible; a representative case of 
them is given in panel (3).

Due to the disk formation processes modifying the initial halo,
however, the comparison made through the dashed curves is far from being
a fair one.
More representative of a hypothetical reality of NFW profiles is the comparison
shown by the continuous curves and the data points. In this case we take the
unique CDM NFW profile which after disk formation has the required asymptotic
velocity of the observed galaxies. In this case we see that whilst for 
F568-1 in panel (1) the fit remains good, in the other two cases no agreement
is possible, and the NFW option must be rejected.

In Fig. 2 we repeat the analysis of Fig. 1, but starting from
a non-singular isothermal halo taken as a King sphere with a form parameter
that will be determined by the best fit to the observations. The data points
remain as in the previous case. The thick continuous curves give the model that
after disk formation has the required asymptotic velocity and the proper 
shape. As it can 
be seen, the entire sample in this case is quite accurately fitted by the
non-singular dark halo taken. The thin continuous curves give the 
contribution to the final
rotation curve due to the dark halo, 
after being modified by the disk formation.
This shows that the final dark halo does indeed dominate 
the resulting rotation curve,
with the exponential disk giving only a minor contribution. The dashed lines
give the rotation curve of the initial dark halo, before disk formation 
took place. This last is given to show the effects of the gravitational pull
suffered by the dark haloes as a result of the concentration of the baryons
during disk formation. A comparison of the dashed thin  
curves and then
of the thin continuous and thick ones, shows that  the two effects which 
process an initial halo into a final rotation curve are sufficient to alter
the conclusions one draws about the existence or not of soft cores. A very
naive direct comparison of a NFW model to the final rotation 
curves might mislead
to an acceptance of that profile; a more detailed look at the problem 
confirms the necessity of initial dark haloes having constant density cores.
For both the Fig. 1 and 2 fits,  we have found values for the baryon
fraction of $0.03<F<0.06$ in which
range our conclusions remain unaffected; the resulting King form 
parameter is $\approx 8$.

For the cluster CL0024+1654 (Tyson et al. 1998) $Rc$ and $\rho_c$ 
have been computed  
transforming the non-singular isothermal density profile to a 
surface density distribution, and then fitting this distribution to
the one deduced from lensing analysis. We have obtained the following 
value for the core radius: $R_c = 100 \pm 20$ kpc. 
Concerning the central
density, we have derived $\rho_c=0.03 \ M_{\odot}$/pc$^3$, but taking into 
account that strong lensing could overestimate the real value of the central 
density by up to a factor $3$ (Wu et al. 1998), we take 
$\rho_c=0.02 \pm 0.01 \ M_{\odot}$/pc$^3$. 
The value of the maximum circular velocity is estimated assuming  
 $V_{m}\approx (GM_{\rm vir}/R_{\rm vir})^{1/2}$, where $R_{\rm vir}$
is a measure of the extent of the virialized region defined as the 
radius within which the average density is $200$ times the background
density, and  $M_{\rm vir}$ is the virial mass within $R_{\rm vir}$.
We use $M_{\rm vir}=4 \ 10^{15} \ M_{\odot}$ and $R_{\rm vir}=3$ Mpc 
(Bonnet et al. 1994). For the galaxy cluster
CL0016+16  we have only an estimate of the central density with large 
error bars. It is therefore difficult to provide an estimate of the 
core radius owing to the large observational uncertainties.

In Fig. 3 we plot $R_c$ vs. $V_{m}$ for our sample of 
dwarfs (filled squares), LSB galaxies (empty squares), and the 
CL0024+1654 cluster (filled circle). 
 Although with some scatter, Fig. 3  
shows that $R_c$ scales as $V_m$.  
In Fig. 4, $\rho_c$ vs. $V_{m}$ is plotted for the same objects 
of Fig. 3. The arrow shows a 
fiducial value derived for a sample of LSB galaxies (de Block 
$\&$ McGaugh 1997). 
As can be clearly seen from
Fig. 4, the halo central density does not correlate with the 
halo size. Although with a large dispersion, one can say that the 
central density of the haloes falls close to $0.02 \ M_{\odot}$/pc$^3$. 

The core radius and central density obtained after a careful
decontamination process of Swaters et al. (2000) data (big empty squares)
are again in
the range of our estimates described previously for the larger body of data,
 showing in fact a somewhat reduced dispersion. This leads
us to expect that if sufficient information were available on
the systems of our larger data set, a more detailed reconstruction 
of their parameters would similarly lead to a reduced scatter
and not show any systematic offsets.

\begin{table}
\caption{The Sample of LSBs, dwarfs $\&$ clusters}
\vspace{0.3cm}           
\begin{tabular}{c|c|c|c|c}  \hline
 &  $V_m$ (km/s) &  
 $R_c$ (kpc) & $\rho_c$ ($M_{\odot}$/pc$^3$) & Ref.   \\ \hline \hline

DDO105  & 88  & 7.5  & $ 0.012$ & (3)  \\
DDO154  & 44  & 2.5  & 0.028 & (1)  \\
DDO170  & 65  & 5  & $ 0.014$ & (2) \\
NGC3109 & 70  & 5.7  & 0.014 & (4) \\
NGC5585 & 85  & 2.8  & 0.08 & (6) \\ 
IC2574  & 70  & 6.5  & 0.01  & (5) \\ \hline

F563-v2 & 110 & 4.52 & 0.034 & (8) \\
F568-1  & 131  & 4.81  & 0.04 & (8) \\
F568-3  & 100  & 6.4  &  0.016 & (8) \\   
F568-v1 & 118  & 5.66   & 0.025 & (8) \\
F571-8  & 135  & 7  & 0.024 & (7) \\
F574-1  & 103  & 4.92  & 0.026 & (8) \\
F583-1  & 85  & 6.8  & 0.01 & (7)   \\
F583-4  & 70  & 7  & 0.007 &  (7)   \\
UGC5999 & 155  & 7.4  & 0.024 &  (7)   \\ \hline

531-G22 & 190 & 9  & 0.03 & (9) \\
533-G4 & 170 & 7.3 & 0.04 & (9) \\ \hline

CL0024+1654 & 2400 & $100 \pm 20$  & $0.02 \pm 0.01$ & (10)\\
\hline

\hline
\end{tabular}\\
 (1) Carignan $\&$ Purton  1998; (2)  Lake  et al. 1990; \\
(3) quoted by Moore 1994; (4) Jobin $\&$ Carignan 1990; \\
(5) Martimbeau  et al. 1994; (6) Cote et al. 1991; \\
(7) de Block $\&$ McGaugh 1997; (8) Swaters et al. 2000;\\
(9) Borriello $\&$ Salucci 2000;(10) Tyson  et al. 1998.\\
\end{table}

Fig. 1-4 show basic properties
of the nature of dark haloes, 
albeit the scatter seen, which theory must address. 
In a previous work  we have  shown that a possible physical mechanism
capable of yielding the observed scale invariance 
for the  halo central density (Fig. 4) is  the
self-interaction of the cold dark particles 
(Firmani et al. 2000). Next section 
is dedicated to a review of other processes that have been suggested for 
producing shallow cores, which however, can be ruled
out because of their disagreement with the scale invariant
behaviour of the halo central density.

\section{Collisionless CDM models}

The hierarchical CDM scenario
of structure formation successfully explains most of the main properties
of the local and high-redshift universe. However, this scenario does not
allow the presence of soft cores in the dark  haloes, a point of
conflict with the observations. Here we explore some general 
alternatives for solving this conflict, without going much 
into the detailed physical mechanisms.

A cut-off in the power spectrum of fluctuations
at some scale, which avoids the formation of small dense substructures
could be expected to yield soft cores.
Such a cut-off appears naturally in warm dark matter scenarios where 
dark matter particles  
remain relativistic until relatively late times.

In this section our goal is to show that a {\it manipulation} of the
power spectrum of the 
primeval density fluctuation field fails in producing haloes
with both soft cores and nearly constant central densities, over the 
entire mass sampled in Fig.4.

Recently Moore and co-workers (1999) using N-body techniques have 
simulated CDM haloes taking a lower cut-off in
the primeval power spectrum of the fluctuation field
as in a warm dark matter universe.
In this case virialized haloes with steep density profiles are formed
with an asymptotic central slope $\rho \propto r^{-1.4}$. 
From our point of view this result 
is basically related to the lack of  particle angular momentum.
In a similar vein, Huss and co-workers (1999) have investigated several
structure formation scenarios ranging from hierarchical 
clustering to spherical radial collapse as a function
of the velocity dispersion, using N-body simulations. 
For steep initial overdensity profiles they never find
virialized haloes with a shallow core. However, for spherical
collapse  with presence of high  velocity
 dispersion, this authors find a final
profile which is slightly shallower than NFW.
Avila-Reese and co-workers (1998) studied the effects of a lower
cut-off in the power spectrum with the addition of some thermal
energy. They find a NFW profile with a central shallow core,
however the scaling properties of the model do not agree
with the observations of dwarf and LSB galaxies.

\begin{figure}
\epsfig{file=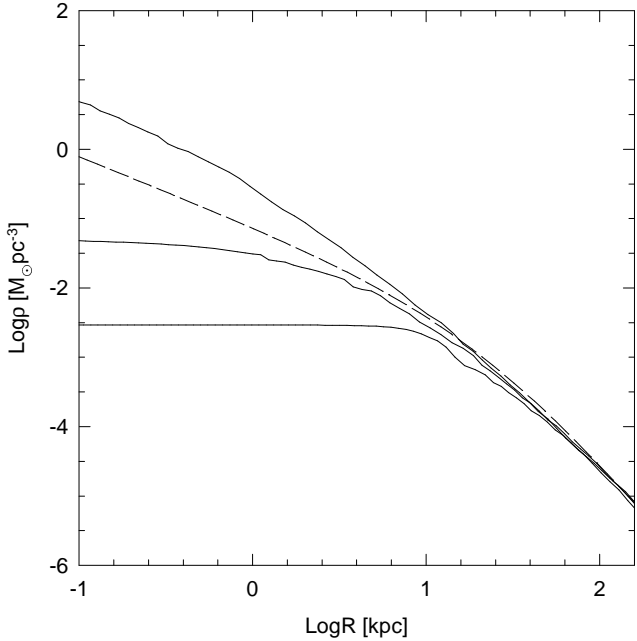,angle=0,width=\hsize,bbllx=79pt,bblly=225pt,
bburx=492pt, bbury=663pt, clip=} 
\@ 
\caption{Dark  matter density profile for a halo of $M= 10 ^{12} M_{\odot}$ and
a cut off in the initial spectrum at $3 \ 10^{10} M_{\odot}$, for
 $\epsilon =0.05;0.27;0.5$ for the upper,middle and lower
curve, respectively; the dashed line is the NFW profile.}
\end{figure}

We stress  that 
 a particle tangential velocity dispersion component (angular 
momentum) could  play a key 
role in producing soft cores.
Hence, we have analyzed extensively
the question investigating the implications
of a lower cut-off in the power spectrum of the cosmic fluctuations 
with the  the contribution of some tangential velocity dispersion. 
We have performed a 
quantitative study with dark matter profiles obtained
by using a semi-numerical
method (Avila-Reese et al. 1998, for details)
that allows us the calculation of spherically symmetric virialized 
structures starting from an arbitrary mass aggregation history (MAH). The
virialized profile of the dark halo is obtained from a generalized 
secondary
infall model including {\it non-radial} motions for the dark matter particles
(tangential velocity dispersion) and taking into account dynamical 
matter redistributions 
through an adiabatic invariance hypothesis. The only free-parameter
of the method is the orbital perihelion 
to aphelion ratio $\epsilon=r_{peri}/r_{apo}$  which is a measure 
of the thermal energy related to the tangential velocity dispersion
of particles. The parameter $\epsilon$ is rather important in 
establishing the central density profile, since it determines 
the approaching of each particle to the center. 
 We set this orbital parameter to the same value provided
by N-body simulations: $\epsilon \approx 0.2-0.3$, and for a CDM model 
we obtain
mass distribution profiles of
virialized structures in agreement with results of numerical 
simulations  (see also Ghigna et al. 1998).

Fig. 5 shows the density profile of a
halo assuming $M=10^{12} \ M_{\odot}$ obtained with 
 a cut-off in the power spectrum for masses lower than
 $3 \ 10^{10} \ M_{\odot}$ and 
 $\epsilon =0.05;0.27;0.5$ for the upper,the middle and lower
curve, respectively; the dashed line is the NFW profile. 
It is interesting to note that when $\epsilon$
increases, a soft core appears.
With decreasing $\epsilon$ giving a more radial
motion for the halo particles, the density profile appears to converge 
 to Moore et al.'s result.

In Fig. 4 we show  the central densities of haloes
as a function of their maximum rotation velocities, dotted curves,
as predicted by the halo structure resulting from a cut-off 
in the power spectrum 
at $3 \ 10^9 \ M_{\odot}$ (upper dotted curve) 
and at $3 \ 10^{10} \ M_{\odot}$
(lower dotted curve) both with $\epsilon=0.27$ . As shown by the plot, only the soft cores of galactic
haloes (LSBs and dwarfs) are reproduced by the model, with 
the best fit being for a very
high  mass cut-off at $3 \ 10^{10} \ M_{\odot}$. 
The model with the value of $\epsilon$ fixed above, does not predict 
a scale invariant central density.
 One might propose an alternative model where $\epsilon$ increases
with mass, however, haloes with masses much larger than the cut-off mass 
assemble most of their mass hierarchically and for this case, even
with $\epsilon \rightarrow 1$, the formation of significant soft cores is not
predicted. On the other hand, the value of $\epsilon=0.27$, which in 
the hierarchical case is accounted by the inhomogeneous collapse
is not easy to justify for haloes formed monolithically.
Although in a warm dark matter scenario particles have a residual 
thermal velocity dispersion, these velocity are not enough 
large as to produce $\epsilon \approx 0.27$. Summarizing, in a 
warm dark matter scenario, haloes near the cut-off scale can be 
shallower than the CDM ones, but for larger haloes the invariance of the
central density with mass is not predicted. This conclusion was 
recently confirmed using  cosmological
N-body simulations (Col\`{\i}n, Avila-Reese $\&$ Valenzuela 2000).

Exploring  an alternative approach we have 
computed halo density profiles truncating the hierarchical halo
mass aggregation histories at a fixed formation redshift, $z_{f}$.
In this process we assume that the halo mass fraction
accumulated up to $z_{f}$ in a CDM mass aggregation history actually
collapses at $z_{f}$. 
 In this 
monolithic  collapse we assume the same orbital parameter as above,
$\epsilon=0.27$,  
while the residual mass is aggregated at the
normal hierarchical rate. 
The aim of this experiment is to show that there is a correlation between the
observed central density and core formation redshift of haloes.
In Fig. 4 we show the halo central densities as a 
function of the maximum circular velocity as derived by a mass aggregation
history starting from $z=5$ (upper dashed curve) and   $z=4$
(lower  dashed curve) and evolving until now with the same average
aggregation history derived from the hierarchical scenario. 
The result 
shows a surprising coincidence between this model and observations.
In light of this result, a central scale invariant dark density 
could be explained if both galaxy and cluster dark cores
collapsed at $z=5$. It is interesting to note that most of distant
observed QSOs and galaxies are at this redshift.   
Although naively attractive, a
physical process that allowed the monolithic collapse of structures 
 at $z=5$ and not before or after this epoch scarcely appears plausible.

Finally, Hogan \& Dalcanton (2000) proposed the existence of a 
limiting phase-space density for dark matter, which
would have been fixed at the epoch of creation. They explore
extensively the cosmological origins and implications for this
assumption, and conclude that it can be accepted or rejected
depending upon whether observations confirm the halo scalings
this hypothesis implies.
On purely
dimensional arguments we can estimate the scalings the maximum volume
density of dark matter would show in this situation as follows:
the hypothesis of a limiting phase -space density,

$$
Q=(\Delta q)^{-3} (\Delta p)^{-3} =const.
$$

yields,

$$
(\rho_{max}/ m_x)(m_x \ V_{m})^{-3}=const.
$$

where $m_x$ is the dark matter particle mass. At constant
$m_x$ we immediately obtain,
$$
\rho_{max}\propto V_{m}^3,
$$
which is the scaling property derived by Hogan \& Dalcanton (2000) for this
assumption. An identification of $\rho_{max}$ with the core
density of dark matter haloes implies a steep scaling of
$\rho_{c}$ with $V_{m}$, again in conflict with the observational
data shown in Fig. 4. An agreement of this approach  with the 
observations requires $Q \propto V_m^{-3}$.

\section{Collisional CDM models}

The success of CDM models at explaining most properties of 
the universe on large scale suggests a viable solution
to the core problem assuming a self-interacting CDM
without dissipative properties as recently proposed 
by Spergel $\&$ Steinhardt (1999).
The core-less NFW density profile,  result of the hierarchical 
process of accretion of haloes via merging of smaller structures,  
shows a velocity dispersion (temperature) that increases outwards
(Cole \& Lacey 1996; Fukushige $\&$ Makino 1997).
Consequently, supposing that CDM particles are 
self-interacting, heat transfer towards the central region triggers
a thermalization process in the dark haloes, avoiding the formation
of a cuspy NFW profile. Since gravitational systems have a negative 
specific heat capacity, heat transfer inwards leads to a 
cooling of the core, amplifying the temperature gradient.
Thus, the heat transfer inwards increases,  
making this a runaway process. The expansion of the core is based on
this physical process if dark matter particles are collisional.
This mechanism is similar to the post-collapse gravothermal
instability 
that characterizes the dynamic evolution of the core
in globular clusters (Bettwieser \& Sugimoto 1984). 
During this process a minimum central density is reached 
after a thermalization time.

The core expansion 
 may halt because particle-particle interactions 
become negligible once the core density decreases 
in such a way that thermal conduction no longer applies. On the 
other hand, if self-interaction is stronger, then an isothermal state 
could be established along the entire halo. In one more extreme case 
the core collapse 
phase (gravothermal catastrophe) is possible. 
Once the velocity 
distribution of the particles reaches a constant velocity dispersion 
under the effects of collisions the 
result is a central non-singular isothermal density profile. 
Any further tendency towards the core gravothermal catastrophe may
be avoided by the competition between: 1) mass and energy aggregation
determined by the halo merging history and 2) a thermalization by 
self-interaction. The mass and the energy aggregation tends to recover 
a NFW density profile (with the corresponding heat transfer inwards) 
while the self-interacting process tries to establish thermal equilibrium.

We argue that the  thermalization process could be the physical
mechanism to explain the presence of shallow cores in dark haloes.
The goal of this work is to analyze   
the final density profiles of virialized haloes in two opposite
cases: 1) when a {\it global} thermal equilibrium is 
reached resulting in a totally thermalized halo well-fitted by a  
King model; 2) when the self-interaction is less effective 
leading to a {\it local} thermal equilibrium.
These cases are two opposite situations in which dark matter 
self-interaction may generate non-singular isothermal cores 
compatible with observations. The first case requires a self-interaction 
cross section sufficiently large in order to reach a global
thermal equilibrium while in the second, a minimal cross-section is necessary 
to induce the thermal equilibrium only
in the core region. 

The possibility that a 
stronger self-interaction is working has to be excluded. Such extreme assumption may
lead the core to the gravothermal catastrophe making the cusp
problem worse.
The case of very strong cross section was studied by Moore and
co-workers (2000) and Yoshida and co-workers (2000) making
use of N-body simulations and treating the dark matter as a 
hydrodynamical fluid. Both works show that in these limits dark haloes
would develop singular isothermal density profiles which are
not in agreement with the observations.

\subsection{King model scaled to CDM haloes}

A global thermal equilibrium is reached when  the self-interaction
cross section is large enough in order for the characteristic time scale 
of collisions across the overall halo to be shorter than the halo
lifetime. This is the case of a very efficient self-interaction
that in a Hubble time leads halo particles to assume a Maxwellian  
velocity distribution and to a virialized  halo density profile well
fitted by a  King model.

We have used the King profile in a cosmological context.
For a given halo mass, we use the King sphere model with the 
corresponding 
total potential 
energy predicted by the mass aggregation
history of the hierarchical theory. In this sense the King
model has to be considered {\it cosmological}. 
 Being the King model depending on three parameters: $\rho_c$,
$r_c$ and P, and since the mass and the potential energy of the halo 
introduce two conditions, one parameter can be freely fixed, e.g. 
$\rho_c$.

According to Fig. 4,
we assume that $\rho_c$ does not depend on mass. So, given the mass,
the potential energy and the central density, the core radius is
univocally determined. In Fig.6 the dotted curves overlaid to the 
observations represent    
  core radii obtained adopting a King profile   with the 
central density fixed to values: 
$\rho_c =0.005 \ M_{\odot}$/pc$^3$
(upper curve);  $\rho_c =0.02 \ M_{\odot}$/pc$^3$ (middle curve)
and  $\rho_c =0.08  \ M_{\odot}$/pc$^3$ (lower curve).
 Vertical dotted lines are the virial masses increasing by a factor
of ten from 
$M=10^{10}$ to $10^{16} \ M_{\odot}$
from left to right.
\begin{figure}
\epsfig{file=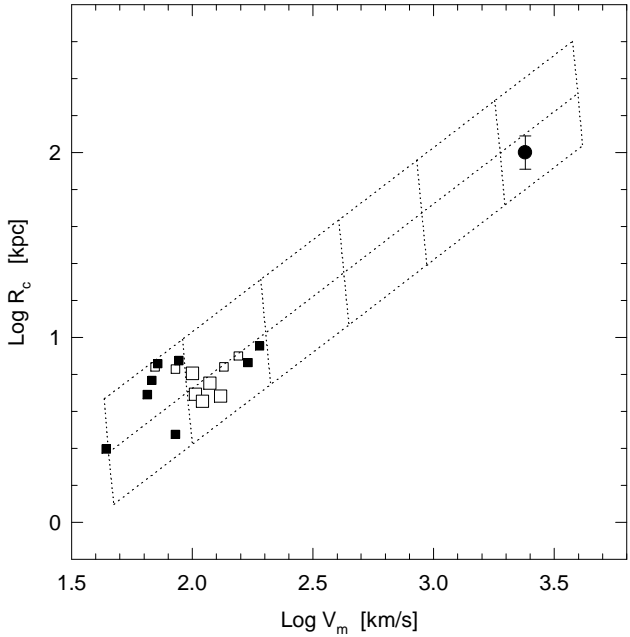,angle=0,width=\hsize,bbllx=79pt,bblly=225pt,
bburx=492pt, bbury=663pt, clip=} 
\@ 
\caption{The core radius vs. maximum circular velocity. 
 Symbols are the same as in Fig. 3.
The dotted lines are curves obtained using a King model with
$\rho_c =0.005;0.02;0.08 \ M_{\odot}$/pc$^3$ from the upper to the 
lower curve.
Vertical dotted lines are for virial masses $M=10^{10}$ to $10^{16} \
M_{\odot}$ 
increasing by a factor ten from left to right.}
\end{figure}
In that figure one sees that a good
agreement with observations is obtained for the scaled King 
models with $\rho_c =0.02 M_{\odot}$/pc$^3$.
The figure shows that the King model is successful in reproducing
halo core radii for the clusters of galaxies; on galactic
scales the coincidence between predictions and the observed trend
seems even better. 

For each scaled King model we have computed the average 
form parameter $P \approx 7, 8.5$, and 9 
for the upper, middle and lower curves in Fig. 6, respectively. 
This detail seems 
interesting because the case of maximum entropy for a King profile
corresponds to a value for the  form parameter $P=8.5$, just
when the halo central density is close to the value of 
$\rho_c \approx 0.02 \ M_{\odot}$/pc$^3$.  
From this point of view one  could conclude
that most galaxies and galaxy cluster are close to  the state of    
maximum entropy.
 
In this sense, the King model is 
capable of predicting the halo central density scale invariance. 
\begin{figure}
\epsfig{file=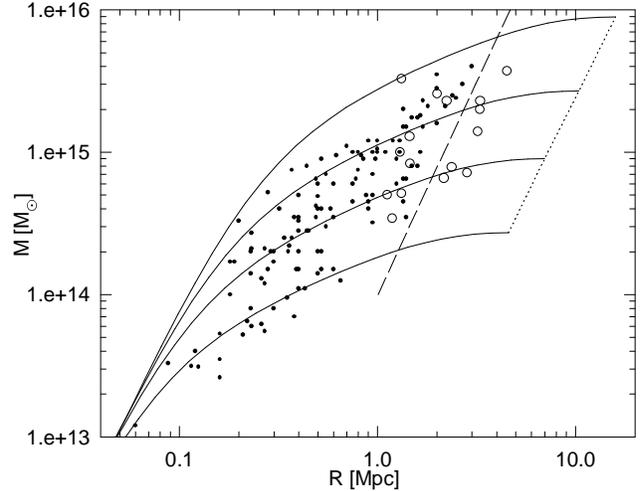,angle=0,width=\hsize,bbllx=81pt,bblly=282pt,
bburx=510pt, bbury=620pt, clip=} 
\@ 
\caption{Solid lines are the expected mass profiles adopting a cosmological
King profile (see text) for haloes of $M_{200}=3 \ 10^{14}$ (lower line);
$10^{15};3 \ 10^{15};10^{16} \ M_{\odot}$ (upper line) and P=8.5.
Points are mass profiles of galaxy clusters 
from weak lensing data (Wu et al.
1998). The open circles represent the clusters from Carlberg (1990). 
The long-dashed line is the relation between 
$M_{vir}$ and $R_{vir}$ in a $\Lambda CDM$ universe.
The dotted line shows the relation between $M_{vir}$ and King 
tidal radius.}
\end{figure}
The expected halo central density
  behaviour is showed with the point-dashed line in Fig.4 . 

Always with the purpose of testing the prediction capabilities of the 
cosmological King model, we have investigated
how realistic is a King configuration 
for the dark mass distribution in galaxy clusters. 
We have compared mass profiles for clusters of galaxies
derived by weak lensing data to the 
predictions of the King model, with the same mass and energy
provided by the hierarchical halo aggregation history and P=8.5
In Fig.7, the dark mass distributions
expected  in a  $\Lambda CDM$-dominated 
universe  are shown with solid lines while  points are for  
 mass profiles of  
29 clusters derived from a published sample 
(Wu  et al. 1998). With solid lines we show  
the predicted mass profiles for haloes
with the following virial masses: $M_{vir}=3 \ 10^{14}$ (lower line);$ 
10^{15};3 \ 10^{15};10^{16} \ M_{\odot}$ (upper line).
No relevant discrepancy appears for the mass profile of clusters, in
particular for objects with a moderate dispersion velocity. 
This result is in agreement with the analysis of the outer part of galaxy
clusters made using lensing techniques as inferred by several authors,
inspite of  their having used the singular isothermal sphere model 
(Williams et al. 1999; Clowe et al. 2000).
The correlation between $R_{vir}$ and $M_{vir}$ 
derived within the  hierarchical picture is plotted in Fig.7 with a
 long-dashed line. Obviously, the criterion used to define the 
virial radius for a cosmological King profile is the same as in the 
hierarchical framework. The open circles drawn in the figure 
represent virial masses of observed clusters as inferred by Carlberg (1996),
while King radii (dotted curve) are indicated to complete the picture.

The crucial shortcoming of this model is that if a thermal equilibrium 
is reached over the entire halo, we should expect that haloes will
be spherical for an extensive region. This prediction strongly disagrees 
with observations of anisotropies for dark mass distribution of the 
cluster MS21137-23, over nearly $100$ kpc, as argued by Miralda-Escud\'{e}
(2000).

In summary, a cosmological King mass profile as an idealized model
to represent a thermal equilibrium extended to the overall halo appears
promising in predicting a thermal equilibrium of maximum
entropy and  producing halo central densities  with a scale invariant
property. Taking into account that a
 strong self-interaction is not needed to generates shallow cores,
 but these may be  produced even when thermalization
is limited to the halo 
inner region. We find it more realistic and likewise
interesting to investigate this alternative possibility.

\subsection{ A non-singular NFW profile}

Of great interest is the case when particle collisions induce 
thermal equilibrium limited to the region of the shallow
cores. This represents the physical situation when a minimal
cross section is given. From a theoretical point of view in this
case, the central
core is thermalized and results in a non-singular isothermal
sphere while in the outer part of the halo the behaviour is provided
by the hierarchical theory of halo merger history represented by
a NFW profile. This is the situation
 in which the thermalization process has a low efficiency and
identifies a region limited by a thermal radius, $R_{th}$. 
This length is not necessarily 
equal to the core radius and as we will see latter may be half as large. 

With the intention of a more quantitative study of this physical
situation we have calculated a profile representative of a {\it
local} thermalization in the inner region of the halo.
We assume a mass density distribution  
 characterized by NFW profile in the outer
part linked to a non-singular isothermal sphere in the core (see Binney
$\&$ Tremaine, table 4-1).
Starting from the hydrostatic equilibrium equation 
the profile is obtained constraining two fitting conditions: 
the continuity of the 
mass density distribution  and of the pressure, in passing from 
the NFW profile to the isothermal sphere. 
The advantage of  such non-singular NFW profile is that it represents 
the most economical approach. 

We have run the model of non-singular NFW profiles in a 
$\Lambda CDM$ universe  
fixing the mass and the central density, hence, we derive 
univocally the core radius
predicted by the model in order to compare it to the observations.
In Fig.8 the dotted curves overlaid to the 
observations represent    
  core radii obtained adopting such a profile   with the 
central density fixed to values: 
$\rho_c =0.005 \ M_{\odot}$/pc$^3$
(upper curve);  $\rho_c =0.02 \ M_{\odot}$/pc$^3$ (middle curve)
and  $\rho_c =0.08  \ M_{\odot}$/pc$^3$ (lower curve).
 Vertical dotted lines are the virial masses increasing by a factor
of ten from 
$M=10^{10}$ to $10^{16} \ M_{\odot}$
from left to right.
 The agreement with the observations appears good.
It is likewise interesting to note that the presence of soft cores as a 
result of weakly self-interacting CDM, is supported by 
Burkert (2000) using a numerical Monte-Carlo N-body
simulation. In this case a Hernquist profile is assumed as initial
condition for the virialized halo and it is shown that in a Hubble time
self-interaction leads to the formation of a non-singular 
isothermal core.
The difference between the
two approaches is that in our model the system is not assumed isolated.
The isothermal soft core is surrounded
by a NFW density profile in the outer part of the halo where mass and thermal
energy are continuously injected as a result of the hierarchical halo 
mass aggregation history. The 
halo characterized by the NFW profile acts as a renewing {\it thermal bath}
feeding the core region. 

With the same procedure used in the previous section we have investigated 
whether the mass profiles
derived by weak lensing data in the outer part of galaxy cluster haloes
may agree with  the predictions of this non-singular NFW profile in  
a hierarchical scenario. Computing the mass profiles for non-singular 
NFW haloes in a  $\Lambda CDM$-dominated universe with $\rho_c=0.02 
M_{\odot}$/pc$^3$,   
 Fig.9 compares directly 
the predictions of the model (solid lines) to the mass profiles for 
29 clusters derived by weak lensing in a published sample 
 (points)(Wu et al. 1998). Models are computed
for the same mass range listed above: 
$M_{vir}=3 \ 10^{14};10^{15};3 \ 10^{15};10^{16} \ M_{\odot}$. 
In this case the agreement with the observations appears a little better 
than for the King model analysed in the previous section.
The correlation between $R_{vir}$ and $M_{vir}$ 
derived within the  hierarchical picture is also plotted in 
 Fig.9 as a long-dashed line. A good agreement between the 
theoretical curves and estimates of virial masses of observed clusters
(open circles) as inferred by Carlberg (1996) is clearly seen.

Now, by adopting this model we are able to investigate the thermal 
radius $R_{th}$.
This length is not necessarily 
equal to the core radius. In CL0024+1654 we predict a core radius roughly of
$R_c=100$  kpc but the 
thermal radius within which the thermalization is working is shorter,
nearly $40$ kpc.
This implies that the core is spherical 
for a few dozen of kiloparsec; beyond this radius the thermalization
does not work and anisotropies will dominate. This is consistent
with Miralda-Escud\'{e} (2000) who finds evidence of
elliptical rather than spherical potentials on scales of the order
of $70$ kpc, as for instance in MS21137-23.

\begin{figure}
\epsfig{file=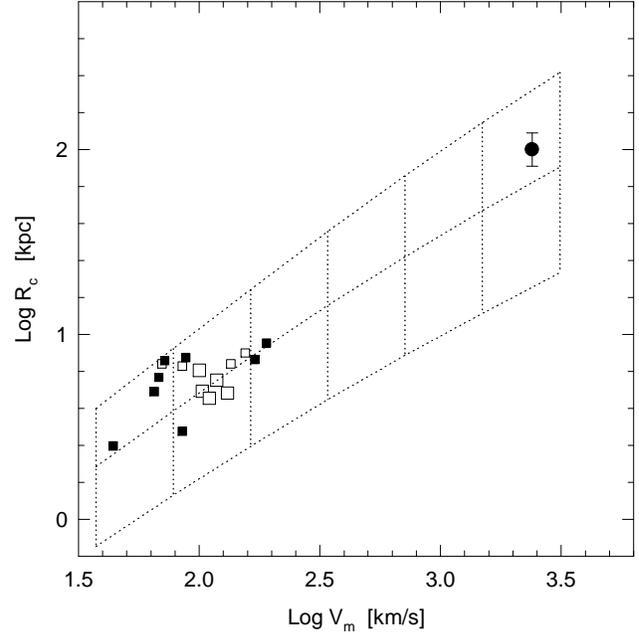,angle=0,width=\hsize,bbllx=79pt,bblly=225pt,
bburx=492pt, bbury=663pt, clip=} 
\@ 
\caption{The core radius vs. maximum circular velocity. Symbols are the
same as in Fig. 3.
 Dotted lines are curves predicted by a non-singular
NFW model with $\rho_c =0.005;0.02;0.08 \ M_{\odot}$/pc$^3$ 
from the upper to the lower curve.
Vertical dotted lines are the virial masses $M=10^{10}$ to $10^{16} \
M_{\odot}$ 
increasing by a factor ten from left to right.}
\end{figure}

\begin{figure}
\epsfig{file=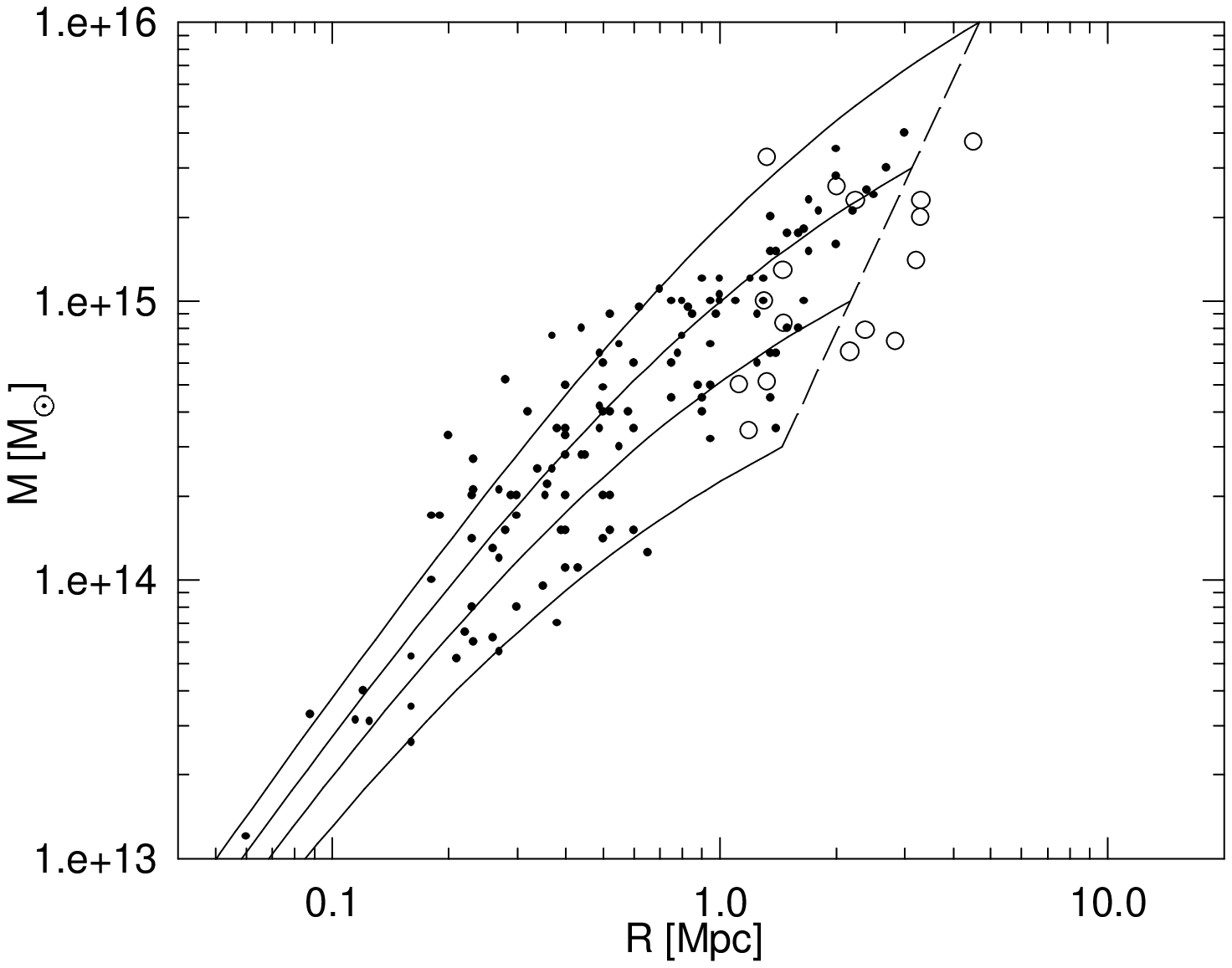,angle=0,width=\hsize,bbllx=81pt,bblly=282pt,
bburx=510pt, bbury=620pt, clip=} 
\@ 
\caption{ Solid lines are the expected mass profiles adopting a non-singular
NFW profile for haloes of $M_{200}=3 \ 10^{14}$ (lower line);
$10^{15};3 \ 10^{15};10^{16} \ M_{\odot}$ (upper line) and 
$\rho_c=0.02 \ M_{\odot}$/pc$^3$.
Points are mass profiles of galaxy clusters 
from weak lensing data (Wu et al.
1998). The open circles represent the clusters from Carlberg (1990). 
The long-dashed line is the relation between 
$M_{vir}$ and $R_{vir}$ in a $\Lambda CDM$ universe.}
\end{figure}

The great advantage offered by assuming a {\it local} thermal equilibrium
is that it allows to estimate the self-interaction cross-section 
directly from the observational data. If $n$ is the 
dark particle number density , $\sigma$ the cross section and $v$ 
the dispersion velocity, assuming in the core a collision time 
$\tau = 1/(n \ \sigma \ v)$ close to the Hubble time, we derive:

\begin{equation}
\frac{\sigma}{m_x} \frac{}{} \approx 4 \ 10^{-25} 
\left( \frac{0.02 \ M_{\odot} \ pc^{-3}}{\rho_c} \right)
\left( \frac{100 \ km \ s^{-1}}{v} \right) \ \frac{cm^2}{GeV}
\end{equation}

with $m_x$ the mass of the dark matter particle and $\rho_c$ the
central density. Even if we ignore the nature of such dark 
matter particles it is important to point out that our model
of self-interaction is characterized by a  cross
section that depends on the halo dispersion velocity i.e., 
as in other classical physical interactions the cross section is a function
of the particle energy. 
 It is important to point out that if one assumes $\sigma/m_x$
 constant  (or equivalently $\rho_c \ v \ t \approx$ constant), 
then the 
halo core radius will scale with v as 
$R_c \propto v^{3/2}$, as proposed by Miralda-Escud\'{e} (2000).
This predicted trend is not favored by the present observational
picture, even though some uncertainty is present. Assuming  $\sigma/ m_x$
as a diminishing function of the dispersion velocity allows the 
self-interaction theory to predict: $R_c \propto v$, in agreement 
with observations.

 During the refereeing of this paper, Yoshida et al. (2000) and 
Dav\'{e} et al. (2000) have carried out cosmological N-body simulations
for a self-interacting CDM model with a constant cross-section.
In a more recent study (Firmani, D'Onghia $\&$ Chincarini 2000), 
using an original cosmological code
based on the collisional Boltzmann equation, we have confirmed
that self-interaction creates in the halo a non-singular isothermal 
core, the outer region remaining as a NFW profile. Assuming 
$\sigma \propto 1/v$ we have proved that a nearly central density 
scale invariance is reproduced. Further, a study of Wyithe et al.
(2000) appeared strengthening the idea of a cross-section inversely
proportional to the collisional velocity.

We wish to point out that when self-interaction induces the local thermal 
equilibrium with the cross-section given above, 
one of the implications of the model is a core formation more 
efficient on galactic than on cluster scales.  
\begin{figure}
\epsfig{file=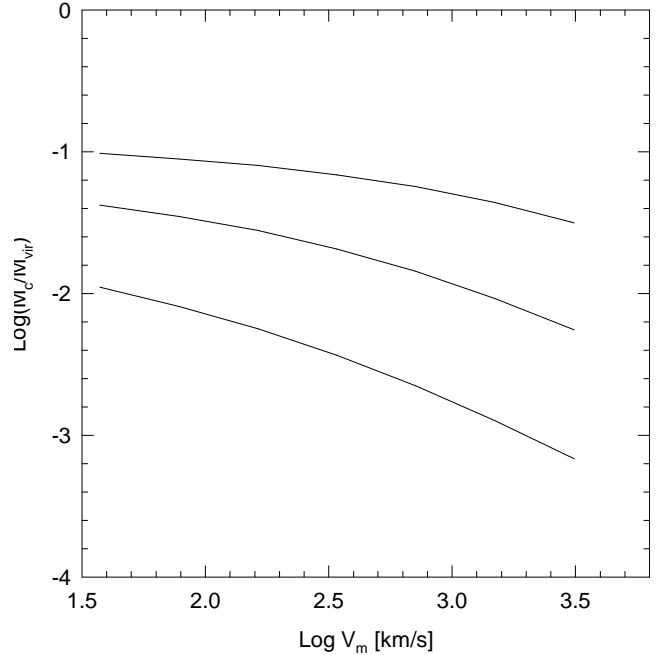,angle=0,width=\hsize,bbllx=83pt,bblly=165pt,
bburx=487pt, bbury=585pt, clip=} 
\@ 
\caption{The expected core mass fraction as a function
of maximum velocity is 
 derived adopting a non-singular NFW profile with the central 
density fixed to values: $\rho_c=0.005 \ M_{\odot}$/pc$^3$ (upper curve),
$\rho_c=0.02 \ M_{\odot}$/pc$^3$ (middle curve) and
$\rho_c=0.08 \ M_{\odot}$/pc$^3$ (lower curve).}
\end{figure}
In Fig.10 we show the expected core mass 
fractions ($M_{c}/M_{vir}$) as a function
of the maximum circular velocity 
 derived adopting the profile with the central density fixed
to values: $\rho_c=0.005 \ M_{\odot}$/pc$^3$ (upper curve),
$\rho_c=0.02 \ M_{\odot}$/pc$^3$ (middle curve) and 
$\rho_c=0.08 \ M_{\odot}$/pc$^3$ (lower curve). 
As shown by the plot, the core mass fraction decreases for more massive haloes
as an intrinsic property of the model.
In light of the cross section value inferred by observations, it 
could be interesting to explore the abundance of galactic subhaloes
making use of numerical simulations and adopting
a weak self-interacting cross section in order to investigate
the role of the ram pressure in disrupting
satellites in dark haloes.

\section{Conclusions}

We have examined 
the observed rotation curves of dark matter dominated
dwarf and LSB galaxies and included the mass density profiles  
of two clusters of galaxies lacking the gravitational contribution
of a central cD.
All these objects show density profiles with evidence of  soft cores.
Assuming an isothermal sphere model, we have
found two scaling properties for haloes from galactic to cluster sizes: 
a linear increase of  
the core radius with increasing maximum rotation velocity 
of the haloes, and a halo central density which remains invariant 
with respect to the mass. 
The difficulty in explaining the observed scaling trends of soft 
cores (in particular a scale invariant central density)  
by a manipulation of the primeval density fluctuation field
led us to assume a self-interaction process as an attractive
physical solution to the core problem. 
We have investigated two opposite physical situations corresponding to a 
{\it global} and a {\it local} thermal equilibrium, 
respectively. These options
are two limiting cases between which dark matter self-interaction
may generate soft cores compatible with the observations.
In particular, a global thermal equilibrium for haloes, reached for
a self-interaction cross section sufficiently large,  
is capable of predicting a scale invariant central density
and  would indicate that the observed haloes are approaching 
the stable equilibrium state of maximum
entropy. Unfortunately, in this model the spherical shape of haloes will be
rather extended, in disagreement with 
lensing data for clusters of galaxies.
 
However, the most attractive case
that we have found is when a local thermal 
equilibrium is induced only in the halo inner region.
This model yields
a good agreement between the observed core radii
 and the theoretical predictions over the entire range sampled.
In light of this
coincidence with the observational data it was possible to derive
{\it inductively} an estimate for the self-interaction cross 
section. 

We stress the importance of confirming the presence of dark halo
soft cores at cluster scales through high resolution mass maps 
using strong lensing 
techniques, which probe directly into the inner regions of galaxy clusters. 
Extending the availability of high-quality galactic rotation curve 
observations in the dwarf and LSB region remains equally important.

\section{Acknowledgments}
We are grateful to S. Gelato for stimulating discussions.\\
ED thanks Fondazione CARIPLO for financial support.

\end{document}